# Lorentzian angles and trigonometry including lightlike vectors[⋆]

Rafael D. Sorkin

Raman Research Institute, Bengaluru, Karnataka, India
and
Perimeter Institute, 31 Caroline Street North, Waterloo ON, N2L 2Y5 Canada
and
Department of Physics, Syracuse University, Syracuse, NY 13244-1130, U.S.A.

address for email: rsorkin@perimeterinstitute.ca

## Abstract

We define a concept of Lorentzian angle that works even when one or both of the directions involved is null (lightlike). Such angles play a role in Regge-Calculus, in the boundary- and corner- terms for the gravitational action, and in the Lorentzian Gauss-Bonnet theorem (for which we provide a proof).

Imagine that you have sliced a pizza into several "wedges", and now you want to reassemble them. Imagine also that you have numbered and marked the individual pieces. When you put two consecutive wedges together, their edges will align perfectly without any special effort on your part. Moreover the opening angles of the wedges will add up to $2\pi$ or 360 degrees, no matter where you made the cuts. In fact, the reassembly would have succeeded equally well if the pizza's radius had been infinite: only the opening angle of each wedge matters. These familiar facts could be summarized by saying that addition of angles is well-defined in the Euclidean plane.

---

[⋆] also available at http://arxiv.org/abs/1908.10022

Now imagine that the plane is not Euclidean but Lorentzian. At first it might seem that nothing much has changed. Some of the cuts will now be timelike instead of spacelike, but consecutive wedges will fit together unambiguously just as before, because distances need to match up along the common boundary between adjacent wedges. But there's an exception! The unambiguous matching of each wedge to its neighbor will fail (assuming a pizza of infinite radius) if the line along which the cut was made was null (lightlike). For the distance-matching in that case reduces tautologically to $0 = 0$. In fact, there corresponds to any radial null-line in the Minkowskian plane $\mathbb{M}^2$ a Lorentz transformation that rescales it by an arbitrary factor, and the identification across the adjacent boundaries is therefore ambiguous by such a factor.★ Thanks to this ambiguity the reassembled pizza can differ from the original. When you put the last wedge in place you might find that a lacuna remains, or vice versa that the last wedge overlaps the first one. In this sense opening angles are not necessarily additive in $\mathbb{M}^2$.

The difference from the Euclidean case is of course that a Lorentzian "angle" parameterizes a boost-transformation, not a rotation, and no boost can take one edge of a wedge to the other edge if the latter is null. Angles involving a null edge fail to be additive because they fail to be defined at all! Nevertheless, we can still bring about a unique matching by restoring the rescaling information that was lost when we chose to cut our "Lorentzian pizza" along a null ray. To preserve this information, it suffices to mark both edges at the same point along the cut, and then to require that the marks be opposite each other when the wedges are brought together.

This suggests that, although one cannot define the angle between, for example, a spacelike ray (half-line) and a null ray, one might be able to define the angle between a spacelike ray and a *truncated* null ray (the truncation being equivalent to a marking as illustrated in figure 1). In other words, the angle between two *vectors* in $\mathbb{M}^2$ might be definable even when the angle between the corresponding *rays* is ambiguous because one of them is lightlike. We will see that a definition of this sort is indeed possible, and that when one adopts it, full additivity is achieved.

---

★ What it more precisely means to "join two wedges together" can be explicated as follows. The wedges are to be embedded isometrically into $\mathbb{M}^2$ in such a way that the edges to be joined coincide. The union of the images of the wedges then gives the geometry of the combined wedge.

Such additivity is basic to simplicial gravity (Regge calculus), where spacetime is treated as a simplicial complex $\Sigma$ built from flat simplices. The action $S$, ignoring boundaries, is then a sum over the interior subsimplices $\sigma$ of dimension two (more generally of codimension two), the summand being the product, $A\theta$, where $A$ is the area of $\sigma$ and $\theta$ is a defect-angle that measures the deviation of $\Sigma$ from flatness at $\sigma$ ("conical singularity"). More specifically, $\theta$ represents the difference between the sum of the dihedral angles formed by the 4-simplices that meet at $\sigma$ and the value this sum would have were $\Sigma$ flat at $\sigma$. In making this definition, one is almost literally following the steps described above to reassemble a pizza, with the opening angles in the pizza corresponding to the dihedral angles here. Thus the Regge-action $S$ rests on a general definition of Lorentzian angle (although strictly speaking, the net defect can be defined by parallel transport without actually needing to define separately the individual opening angles.)

The above pertains to 2-simplices $\sigma$ that are interior to $\Sigma$. When $\sigma$ lies on $\partial\Sigma$, the boundary of $\Sigma$, one must replace $\theta$ by the analogous angle that measures the deviation from *extrinsic* flatness of $\partial\Sigma$, viz. the difference between the sum of the dihedral angles formed by the 4-simplices that meet at $\sigma$ and the value this sum would have were $\partial\Sigma$ extrinsically flat at $\sigma$.

One can think of this defect in terms of the dihedral angle between the two boundary 3-simplices that meet at $\sigma$, and when one does so, the so-called *corner terms* needed (in the continuum) to supplement the Einstein-Hilbert action emerge automatically, this being apparently the simplest way to derive them. It is plausible, moreover, that the entire boundary action in the continuum, including even the contribution of the null boundaries, could be interpreted as being the sum of an infinite number of infinitesimal corner terms.

As a simple consequence of these considerations of "Lorentzian trigonometry", we will be able in Section 9 to deduce a Lorentzian Gauss-Bonnet theorem, in probably its most general form.

In what follows, we explore to what extent one can define, in the Lorentzian plane, angles which are finite and add up consistently, even when one or both of the directions involved is lightlike. Taking additivity as our guide, we will begin with a pair of spacelike vectors and progress finally to the various null cases, whose analysis will be the main goal of our investigation.

## 1. Two identities useful in defining and adding angles

In the Euclidean plane the existence of opening-angles which add consistently under juxtaposition of wedges can be traced to the properties of the rotation group $SO(2)$, in particular the fact that no matter how wide or narrow a wedge is, one can always find a rotation taking one of its edges to the other. In $\mathbb{M}^2$ this is in general not possible, as emphasized above. Only in certain cases can one define an opening angle by reference to the group $SO(1,1)$. This is possible for a convex wedge with spacelike edges, but not e.g. for a wedge with one edge spacelike and the other timelike. Is there then a different definition of angle which does generalize successfully to the Lorentzian case?

That there should be one appears when one considers that the additivity of angles can in the end only be a statement about vectors, and that such a statement must ultimately reduce to a relationship among their inner products. The equation $\cos\theta = a \cdot b/|a||b|$ determines the Euclidean angle between (nonzero) vectors $a$ and $b$, and so the fact that $\theta(a,b) + \theta(b,c) = \theta(a,c)$ when $b$ lies between $a$ and $c$ must correspond to an equation that relates $a \cdot c$ to $a \cdot b$ and $b \cdot c$. Modulo a choice of signs and "analytic branches" this equation is clearly $\cos^{-1}(a \cdot c) = \cos^{-1}(a \cdot b) + \cos^{-1}(b \cdot c)$, which can also be written in terms of logarithms and square roots. Underlying this equation is a simple identity whose Lorentzian counterpart will lead us to definition of angle that works for all pairs of vectors, be they timelike, spacelike, or lightlike.

In fact we will need two identities in the Lorentzian case which we now state and prove. To that end let us define, for any two vectors, $a$ and $b$, in $\mathbb{M}^2$,

$$Z(a,b) = a \cdot b + ||a \wedge b|| \tag{1}$$

$$\bar{Z}(a,b) = a \cdot b - ||a \wedge b||$$

where $||a \wedge b||$ is the norm of $a \wedge b$, namely the positive square root of $(a \cdot b)^2 - (a \cdot a)(b \cdot b)$. Notice here that $(a \cdot b)^2 - (a \cdot a)(b \cdot b)$ is always a positive real number because the square of the timelike two-form $a \wedge b$ is on one hand always negative, and on the other hand equal to $(a \cdot a)(b \cdot b) - (a \cdot b)^2$. Our first, trivially proven identity is then

$$Z(a,b) \ \overline{Z}(a,b) = |a|^2 \ |b|^2 \tag{2}$$

where $|v|^2$ is by definition $v \cdot v$, the inner product of $v$ with itself. (We will adopt the Lorentzian signature for which $v \cdot v$ is positive for spacelike $v$ and negative for timelike $v$. Neither of our identities will depend on this choice, however.)

The second, less trivial identity involves three vectors $a$, $b$, $c$, of which the second is between the other two, in the sense that it is a linear combination of the other two with positive coefficients (see figure2):

$$b = \alpha a + \gamma c \qquad \alpha, \gamma \geq 0$$

There holds in this situation a second identity,

$$Z(a,b) \ Z(b,c) = |b|^2 \ Z(a,c) \tag{3}$$

To prove it, let us observe first that since (3) is homogeneous in the vectors involved, we can without loss of generality assume that $\alpha = \gamma = 1$, whence $b = a + c$. From this follows also $a \wedge b = a \wedge (a + c) = a \wedge c$, and similarly $b \wedge c = a \wedge c$, so all three norms $||a \wedge b||$, $||a \wedge c||$, and $||b \wedge c||$ are equal to $||a \wedge c||$. Substituting then $a + c$ for $b$ in (3), using the definition (1), and expanding terms, we arrive at $0 = 0$.

## 2. Opening-angle of a spacelike wedge in the first quadrant

The two null lines through the origin in $\mathbb{M}^2$ divide the plane into four "quadrants" as shown in figure 3. Consider, to begin with, the case where $a$ and $b$ lie in the first quadrant. There is then a unique Lorentzian boost that takes the ray through $a$ to that through $b$, and the corresponding boost-parameter $\theta$ with its standard normalization has, as is well known, the magnitude $\theta = \cosh^{-1} \widehat{a} \cdot \widehat{b}$, where $\widehat{a} = a/|a|$ and similarly for $\widehat{b}$. (One can easily verify this explicitly in null coordinates, $u = x - t$ and $v = t + x$.) Without any loss of generality, we could of course constrain $a$ and $b$ to be normalized in this case, but it will be useful below to have left them general. We therefore write the formula for the Lorentzian angle in the present case as

$$\theta(a,b) = \cosh^{-1} \frac{a \cdot b}{|a||b|} \tag{4}$$

Notice here that we are taking the opening-angle $\theta$ of a spacelike wedge to be real and positive by convention. This is the first of several conventions we will be led to make in defining Lorentzian angles. Notice also that (4) is symmetric in $a$ and $b$, consistent with the fact that we are not tying the definition of angle to any choice of orientation in $\mathbb{M}^2$.

To put (4) into a form better suited to our identities, let us recall that

$$\cosh^{-1} z = \log(z + \sqrt{z^2 - 1}) \ , \tag{5}$$

in virtue of which we can rewrite (4) as

$$\theta(a,b) = \log \frac{Z(a,b)}{|a||b|} \tag{6}$$

In conjunction with the key requirement of additivity, this equation will determine all other angles $\theta(a,b)$ almost uniquely.

## 3. Opening angle of a wedge with one edge in quadrant I and the other in quadrant II

The next case to consider is that of the angle between a spacelike vector $a$ and a timelike vector $b$. For example $a$ could be in the first quadrant and $b$ in the second, as in figure 4.

Recalling that we have adopted additivity of opening-angles as our guiding light, let us observe that with the definition (6), additivity within quadrant I is guaranteed by the identity (3). Since such identities are preserved under analytic continuation, let us continue to employ the analytic form (6) when $b$ moves from the first quadrant to the second. This however will only determine $\theta$ fully when we decide which sheet of the corresponding Riemann surface $\theta$ should lie on, or equivalently which "branches" of 'log' and of $|b| = \sqrt{b \cdot b}$ should be chosen.

Imagine now that $a$ remains fixed, while $b$ moves continuously from $b = a$ to a point in the second quadrant. Then $Z(a, b)$ will remain strictly positive, but $|b|$ will pass through zero when $b$ crosses the lightcone. The ratio $z = Z(a, b)/|a||b|$ will thus trace a path through the point at infinity in the Riemann sphere, where the logarithm has a branch-point. To resolve the resulting ambiguity in $\log z$ we can adopt the "$i\varepsilon$ prescription" of reference [1], where the metric was given a positive-definite imaginary part. In application to the present problem, this simply means that $a \cdot a$ will acquire a positive imaginary part, or equivalently, $a \cdot a \to a \cdot a + i\varepsilon$, meaning that $a \cdot a$ will circle the origin in a positive or anticlockwise direction. In consequence, $z$ will circle the origin in a negative direction, and $\log z$ will pick up an imaginary piece, $-i\pi/2$. When $b$ completes its journey, we will thus have $|b| = \sqrt{b \cdot b} = i\sqrt{|b \cdot b|}$, together with

$$\theta(a, b) = \log \frac{Z(a, b)}{|a|\,|b|} = \log \frac{Z(a, b)}{||a||\,||b||} - i\pi/2 , \tag{7}$$

where $||v||$ denotes the absolute value of $|v|$, and where in the middle expression one should interpret $\log(1/i)$ as $-i\pi/2$.

We can put this into a more familiar form by dividing through by $||a||\,||b||$ and defining $\widehat{b} = b/||b||$ and similarly for $\widehat{a}$. The result is

$$\theta = \log\left(\widehat{a} \cdot \widehat{b} + \sqrt{(\widehat{a} \cdot \widehat{b})^2 + 1}\right) - i\pi/2 ,$$

an equation which can also be written as

$$\theta(a, b) = \sinh^{-1}(\widehat{a} \cdot \widehat{b}) - i\pi/2 \tag{8}$$

when we recall that $\sinh^{-1}(x) = \log(x^2 + \sqrt{x^2 + 1})$. Notice here that the plus sign under the square root came about because $b$ was timelike.

REMARK The steps leading to (8) had the effect of substituting $y = \widehat{a} \cdot \widehat{b}$ in the identity, $\cosh^{-1}(-iy) = \sinh^{-1}(y) - i\pi/2$, which holds for a suitable identification of the Riemann surfaces of $\cosh^{-1}$ and $\sinh^{-1}$ with each other. To prove this identity, recall that $\cos y = \cosh(iy)$, $\sin y = -i\sinh(iy)$, and make these substitutions in the further identity $\sin x = \cos(x - \pi/2)$, taking then $y = \sinh(ix)$.

We have in this section made a second choice of convention in taking $|b|$ to be positive-imaginary rather than negative, and therefore taking the imaginary parts of (7) and (8) to be negative rather than positive. The physical meaning of this sign shows up in connection with topology-changing spacetimes, where the defect angle enters into the gravitational action-functional (see [1]). Additivity alone would not have forced the angle to be complex, but the requirement that the usual formulas of trigonometry continue to hold for Lorentzian metrics does demand it (see [2]), and the Lorentzian Gauss-Bonnet theorem also requires it. Whether it has further significant consequences remains to be seen.

## 4. Opening angle of a timelike wedge in quadrant II

We could obtain $\theta(a, b)$ in this case by starting with (8) and analytically continuing $a$ from quadrant I to quadrant II, but since we can deduce it directly from the spacelike case by invoking additivity, it seems simpler and more instructive to proceed that way.

In figure 5, the pairs of vectors, $a$ cum $b$ and $a'$ cum $b'$ respectively delineate two wedges related by a Lorentz-boost, illustrating the familiar fact that spacelike and timelike vectors rotate in opposite directions under a boost. But since angles are defined by the intrinsic geometry, they are Lorentz-invariant, and $\theta(a, b)$ must equal $\theta(a', b')$. On the other hand, additivity of opening-angles requires that $\theta(a, b) = \theta(a, a') + \theta(a', b') + \theta(b', b)$ from which it follows that $\theta(a, a') + \theta(b, b') = 0$.

In other words, the angle separating two timelike vectors in quadrant II has a magnitude equal to that of the boost relating them, but its sign is negative since it has to be opposite to that of angles within quadrant I: timelike wedges have *negative* opening-angles. In simplicial gravity, this opposite sign is what guarantees that the defect-angle that enters into the Regge action is defined consistently (see [2] and [3]).

Now the boost-angle between two timelike vectors in the same quadrant is given, similarly to (4), by $\cosh\theta = |\widehat{a} \cdot \widehat{b}|$, so we have for this case (and for our choice of signature that makes $\widehat{a} \cdot \widehat{b}$ negative)

$$\theta(a, b) = -cosh^{-1}(-\widehat{a} \cdot \widehat{b}) \;, \tag{9}$$

which can also be written with the aid of (5) and the definition of $\overline{Z}$ as

$$\theta(a, b) = -\log \frac{\overline{Z}(a, b)}{|a|\,|b|} \tag{10}$$

Notice here that because $a$ and $b$ are timelike, $|a|$ and $|b|$ are pure imaginary, and so both $\bar{Z}(a,b)$ and $|a|\,|b|$ are negative.

## 5. Opening angle of a wedge with one null edge

Our formulas derived so far allow one to deduce the opening angle of any wedge $W$ whose edges are either spacelike or timelike but not null. This includes wedges whose edges lie in opposite quadrants, as well as non-convex wedges like that shown in figure 6, and wedges which overlap themselves. In all such cases, it suffices to subdivide $W$ into sub-wedges each of which is convex and fits into one of the cases analyzed above. Additivity will then guarantee that summing the angles of the sub-wedges will produce a value independent of how the subdivision was done.

The only outstanding situation, therefore, is that where either or both edges of $W$ are null (lightlike). One might think that no useful angle could be defined at all then, because the opening angle of, for example, a spacelike wedge diverges as one of its edges approaches the light cone: a boost that could take a non-null direction to a null direction would have to be infinite. It is therefore surprising that — as explained earlier — one can actually define finite and additive opening-angles in this situation if one works with wedges whose null edges are *marked*. That is, one can successfully define an angle between a non-null ray and a null vector, or between two null vectors. And these definitions arise naturally in the context of quantum gravity.

To see how this comes about, consider a triplet of vectors as seen in figure 7. Let $a$ and $b$ lie in quadrant I and $c$ in quadrant II, with $b$ near to, but not actually on, the lightcone that divides the quadrants from each other. We know that for these non-null vectors, $\theta(a,b) + \theta(b,c) = \theta(a,c)$. Let us try to rearrange the latter equation in such a way that it will remain well defined when $b$ approaches the null vector $n$.

To that end, let us call on the identity (3), which expresses additivity of angles in terms of inner products of vectors:

$$Z(a,b)\ Z(b,c) = |b|^2\ Z(a,c) \tag{3}$$

As $b \to n$, $|b|^2 \to 0$, and (3) must trivialize to $0 = 0$. Hence either $Z(a,n)$ or $Z(n,c)$ must vanish, and it is easy to see that (with our signature of $(-+++)$) $Z(n,c)$ does so, because $n \cdot c < 0$, $c$ being timelike. The equation $0 = 0$ is not very useful, of course, but if we multiply through by $\bar{Z}(b,c)$ before taking the limit, the zeros will cancel and what remains will suggest appropriate definitions of $\theta(a,n)$ and $\theta(n,c)$.

Proceeding this way and taking note of (2), we obtain successively

$$Z(a,b)\, Z(b,c)\, \bar{Z}(b,c) = |b|^2\, Z(a,c)\, \bar{Z}(b,c)$$

$$Z(a,b)\, |b|^2\, |c|^2 = |b|^2\, Z(a,c)\, \bar{Z}(b,c)$$

$$Z(a,b)\, |c|^2 = Z(a,c)\, \bar{Z}(b,c) \tag{11}$$

Now let $b \to n$, obtaining in the limit (which is now smooth)

$$Z(a,n)\, |c|^2 = Z(a,c)\, \bar{Z}(n,c)$$

But because $n \cdot n = 0$ and $a \cdot n > 0$, we have firstly

$$Z(a,n) = a \cdot n + \sqrt{(a \cdot n)^2 - (a \cdot a)(n \cdot n)} = 2a \cdot n$$

and secondly (because also $c \cdot n < 0$)

$$\bar{Z}(n,c) = n \cdot c - \sqrt{(n \cdot c)^2 - (n \cdot n)(c \cdot c)} = 2n \cdot c$$

Equation (11) thus becomes

$$Z(a,c) = \frac{2\, a \cdot n}{2\, c \cdot n}\, |c|^2 \tag{12}$$

REMARK  It is obvious geometrically that $a \cdot c$ must be determined by $a \cdot n$ and $n \cdot c$, and indeed, one can show that with $n$ null, this relationship takes the pretty form,

$$2\, a \cdot c = \frac{c \cdot n}{a \cdot n} - \frac{a \cdot n}{c \cdot n}$$

Equation (12) is nothing but a convenient form of this last equality.

In (12) we have the result we need, but it is not quite in the form we need. In order to make contact with $\theta(a,c)$, we can divide through by $|a|\,|c|$, obtaining thereby

$$\frac{Z(a,c)}{|a|\,|c|} = \frac{2\, a \cdot n/|a|}{2\, c \cdot n/|c|}$$

or with reference to equation (7),

$$\theta(a,c) = \log \frac{Z(a,c)}{|a|\,|c|} = \log \frac{2\,a\cdot n/|a|}{2\,c\cdot n/|c|}\,, \tag{13}$$

which since $a\cdot n$ and $|a|$ are both positive real numbers, can be written without any loss of phase information as

$$\theta(a,c) = \log(2\,a\cdot n/|a|) - \log(2\,c\cdot n/|c|) \tag{14}$$

In (14), $\theta(a,c)$ is expressed as a sum of two terms, the first involving only $a$ and $n$, and the second only $b$ and $n$ — a form suited perfectly to angle additivity! The suggestion springing from (14) then, is to define $\theta(a,n)$ to be $\log(2\,a\cdot n/|a|)$ and $\theta(c,n)$ to be $-\log(2\,c\cdot n/|c|)$. (We could of course have dropped the factor of 2 from these formulas, but it turns out that one obtains a more uniform set of angle-definitions by retaining it.)

This is basically the course we will follow, but with one or two amendments. The first problem is that the arguments of the logarithms are not dimensionless.[†] For dimensional consistency it seems necessary to introduce a reference length $\ell_0$, leading to the amended definitions,

$$\theta(a,n) = \log \frac{2\,a\cdot n}{|a|\,\ell_0} \qquad \text{(provisional)} \tag{15a}$$

$$\theta(c,n) = -\log \frac{2\,c\cdot n}{|c|\,\ell_0} \qquad \text{(provisional)} \tag{15b}$$

Here, in the second equation, one is meant to interpret $|c|$ as a positive imaginary number and to interpret $\log i$ as $+i\pi/2$, following the conventions we have been adhering to throughout. If we do so then $\theta(a,n)$ will be real, while $\theta(n,c)$ will, in common with $\theta(a,c)$, have $-i\pi/2$ as its imaginary piece. The angles will thus add consistently as desired:

$$\theta(a,c) = \theta(a,n) + \theta(n,c) \tag{16}$$

---

[†] This statement presupposes that an inner product like $a\cdot n$ has the dimensions of length-squared, as one would normally expect it to do. However if $n$ and/or $a$ had other dimensions than the usual ones, as a normalized vector like $v/||v||$ does for example, then the arguments of the logarithms would not necessarily be dimensionful.

The second problem, or rather ambiguity, is that the provisional pair of definitions (15) is only one among many equally consistent possibilities. Just as we could have omitted the factor of 2 in both (15a) and (15b), we could, without in any way spoiling (16), have added any complex constant $c$ to $\theta(a, n)$ and subtracted it from $\theta(c, n)$. To do justice to this freedom, we should perhaps have included such a constant $c$ explicitly in equations (15). Complicating our equations that way can be avoided, however, if we notice that any such $c$ can be absorbed into $\ell_0$, provided that we are willing to let $\ell_0$ become complex. If we agree to keep in mind that $\ell_0$ might in principle be complex, then we can treat the choice of $c$ as purely a question of notational convenience. (Or still better, could we identify a convincing reason why one value of $c$ should be selected as the "right one"? Perhaps, given that a change of $c$ would modify the "corner terms" in the gravitational action, quantum gravity could provide such a reason, but for the moment, I know of none.)

What then would be the most convenient choice? On one hand, (15) looks simple, on the other hand it introduces a hard-to-remember asymmetry into our angles. Whereas $\theta(a, n)$ in (15a) is purely real, as if $n$ had been displaced infinitesimally into quadrant-I, $\theta(c, n)$ in (15b) has an imaginary piece of $-i\pi/2$, as if $n$ had been displaced infinitesimally into quadrant-II. More symmetrically we could imagine $n$ as falling precisely on the lightcone, exactly midway between quadrant I and quadrant II, in which case we would attribute equal contributions of $-i\pi/4$ to both $\theta(a, n)$ and $\theta(c, n)$. With this convention every wedge with a single lightlike edge will acquire an imaginary contribution of $-i\pi/4$. Neither alternative seems clearly better than the other, but mnemonically the second is perhaps preferable. For definiteness, I will adopt it in the remainder of this paper.

No matter which alternative we choose, the *real* parts of $\theta(a, n)$ and $\theta(c, n)$ will be given by the same expressions:

$$\operatorname{Re}\theta(a, n) = \log \frac{2\,|a \cdot n|}{||a||\,\ell_0} \tag{17a}$$

$$\operatorname{Re}\theta(c, n) = -\log \frac{2\,|c \cdot n|}{||c||\,\ell_0} \tag{17b}$$

With the conventions we have adopted the full formulas in the present case will be as follows.

When $a$ is spacelike and $n$ is null (both in same closed quadrant)

$$\theta(a,n) = \log \frac{2|a \cdot n|}{||a||\,\ell_0} - i\pi/4 \tag{18}$$

When $b$ is timelike and $n$ is null (both in same closed quadrant)

$$\theta(b,n) = -\log \frac{2|b \cdot n|}{||b||\,\ell_0} - i\pi/4 \tag{19}$$

Notice that two more conventions have entered our discussion, one being the choice of reference length $\ell_0$, the other being how to apportion the imaginary contribution $-i\pi/2$ between the "almost spacelike" wedge that lies between $n$ and $a$ and the "almost timelike" wedge that lies between $n$ and $c$.

That we have had to introduce the length $\ell_0$ means that, insofar as null vectors are involved, angles have lost their familiar conformal invariance. Under rescaling of the metric they will now change by a logarithmic additive constant. Perhaps it should not be a surprise that when one manages to convert an angle that a priori would have been infinite into something finite, an additive ambiguity arises. On the other hand, it is also true that the ambiguity cancels in certain sums or differences of angles. (Indeed its cancellation in the sum $\theta(a,n) + \theta(n,b)$ was precisely the requirement that led us to our definitions in the first place.) It cancels in particular if one compares the angle $\theta(a,b)$ between two vectors with the angle between the same two vectors with respect to a different metric, provided that neither $a$ nor $b$ goes from being null to non-null or the reverse. It is this fact which guarantees that the "double path integral" is unambiguously defined in gravity, even for regions with null boundary-portions (cf. [4] [5]). In light of these indications, we might be tempted to declare that only quantities from which the reference length $l_0$ has dropped out are meaningful.

## Heuristic derivation of equation (17a)

Given the somewhat circuitous route we took to reach the definitions (18) and (19), it might be reassuring to see how one could have arrived at the same result by a different path which, though it takes some liberties with logic, is more direct and intuitive. In the following we will be using the familiar fact that, when it acts on a null vector $n$, a boost $\Lambda$ of rapidity-parameter $\theta$ takes $n$ to $\Lambda n = e^\theta n$.

We seek to deduce the angle between the same two vectors as before, $a$ (spacelike) and $n$ (null), assuming for simplicity that $a$ is normalized ($a \cdot a = 1$). Now let $b = \Lambda a$ be the result of applying to $a$ a boost transformation $\Lambda$ of angle $\eta$ which will carry it toward the lightcone, as in figure 8.

Since $\Lambda$ is an isometry, $b$ is also normalized. We know that $b$ can never truly reach the lightcone, but by taking $\eta$ arbitrarily great it can come as "close" as desired. Imagine then, that $b$ has come so close to being lightlike that it is "for all practical purposes" a multiple of $n$, a very great multiple since the boost was so large. In other words $b \approx n'$ for some $\lambda \gg 1$ such that $n' = \lambda n$. (For definiteness, we can determine $n'$ by the condition that $b - n'$ be lightlike.) Because $n$ and $n'$ are both on the lightcone, a unique boost takes $n$ to $n'$; let its parameter be $\gamma$, so that $n' = e^{\gamma} n$. Because Lorentzian angles are taken by definition to be boost-parameters whenever this makes sense, we can say that $\theta(n, n') = \gamma$. Combining this with the approximate equality, $\theta(a, b) \approx \theta(a, n')$, we learn that $\theta(a, n) = \theta(a, n') - \theta(n, n') \approx \theta(a, b) - \theta(n, n') = \eta - \gamma$.

Now let us put this plan into action. Let $u$ and $v$ be null vectors such that $u \cdot v = 1/2$ with $v$ pointing toward the future and $u$ toward the past. Without loss of generality, we can take $a = u + v$ and $n = \alpha v$. Then (for $\eta \gg 1$):
$b = \Lambda a = e^{\eta} v + e^{-\eta} u \approx e^{\eta} v$,
$n' = e^{\eta} v = (e^{\eta}/\alpha) n$,
$\gamma = \log(e^{\eta}/\alpha) = \eta - \log \alpha$,
$\theta(a, n) \approx \eta - \gamma = \log \alpha$.
Now compare this with, $a \cdot n = (u + v) \cdot (\alpha v) = \alpha/2$.
We conclude that
$$\theta = \log \alpha = \log(2(a \cdot n)) \,, \qquad (20)$$
in perfect agreement with (15a).[♭]

---

[♭] Even the factor of 2 is the same! That no dimensionful constant like $l_0$ appears in (20) illustrates the point made in a previous footnote. Because $b \cdot b = a \cdot a = 1$ is a pure number, and because we have assumed $n = b/\alpha$ with $\alpha$ a pure number, the combination $2(a \cdot n)$ is also a pure number, and is already dimensionless without the need for any conversion factor.

## 6. Opening angle of a wedge with two null edges

Consider a wedge $W$ with two lightlike edges marked by vectors $a$ and $b$. If we confine ourselves to convex wedges, there will be three sub-cases to consider according as $W$ fills out a spacelike quadrant, a timelike quadrant, or an entire half-space (figure 9). In all three situations, the required opening angle $\theta(a,b)$ follows uniquely via additivity from the formulas we have already derived. (In addition — if we want to include it — there is a fourth sub-case of an "infinitely thin null wedge", but we will postpone its consideration to the next section.)

Referring to the first situation depicted in the figure, let $a$ and $b$ be null vectors on the boundaries of quadrant I, and let $w$ be a unit vector between them. It is easy to verify (e.g. by introducing orthonormal vectors $\widehat{x}$ and $\widehat{t}$, and taking $w = \widehat{x}$, $a = \lambda(\widehat{x} - \widehat{t})$, $b = \mu(\widehat{x} + \widehat{t})$) that

$$(2a \cdot w)(2w \cdot b) = (2a \cdot b) , \qquad (21)$$

the logarithm of which says (all the factors being positive)

$$\log(2a \cdot w) + \log(2w \cdot b) = \log(2a \cdot b)$$

Comparing with (18) and remembering that $|w| = 1$, we conclude that for $a$ and $b$ both null and in the closure of quadrant I,

$$\theta(a,b) = \log \frac{2|a \cdot b|}{\ell_0^2} - i\pi/2 \qquad (22)$$

This was the case of a spacelike quadrant. In the opposite case of a timelike quadrant the calculation is equally simple, and reveals that for $a$ and $b$ both null and in the closed quadrant II,

$$\theta(a,b) = -\log \frac{2|a \cdot b|}{\ell_0^2} - i\pi/2 , \qquad (23)$$

exactly the same formula as (22) except for the sign of the real part, which as we know will always flip when we go from a spacelike to a timelike wedge.

Turn now to the third situation illustrated in the figure, where $a$ and $b$ are anti-parallel null vectors and the wedge $W$ is a half-space, and take $m$ to be a null vector which lies between $a$ and $b$ within $W$. The calculation is even simpler in this case, since no non-null edges are involved. Starting from $\theta(a,b) = \theta(a,m) + \theta(m,b)$ and substituting the values (22) and (23), we obtain

$$\theta(a,b) = \log \frac{|a \cdot m|}{|b \cdot m|} - i\pi = \log(-a:b) - i\pi \qquad (24)$$

In this last equality, $a:b$ is the ratio of $a$ to $b$, defined as that number $\lambda$ (necessarily negative when $a$ and $b$ are antiparallel) for which $a = \lambda b$.

How does it happen that $a:b$ and not $b:a$ occurs in (24)? At first sight, it might seem that $a$ and $b$ play symmetrical roles, but an examination of the different signs in (22) and (23) reveals the relevant difference: $a$ forms a spacelike wedge together with $m$ whereas the wedge bounded by $b$ and $m$ is timelike. Or to put the distinction another way, $a$ would become spacelike if it were to move into $W$, whereas $b$ would become timelike were it to do so.

The calculations here illustrate the mnemonic that each null edge contributes $-i\pi/4$ to the angle (the simplicity of this rule being an advantage of the conventions we adopted in the previous section.)

## 7. "Opening angle" of a "sliver" with parallel null edges

In Section 5 we observed that the boost-angle between two parallel null vectors $n$ and $n'$ is given by the log of their ratio, i.e. $\log \lambda$ if $n$ and $\lambda n$ are the two vectors in question. But the sign of such an expression is in general ambiguous, unless we can decide whether to form the ratio as $n:n'$ or as $n':n$ (whether the answer should be $\log \lambda$ or $\log \lambda^{-1}$). It might seem that this question was moot because the type of "infinitely thin null wedge" to which it refers would never arise in practise anyway. (After all, why would you want to slice a pizza twice in the same place?) In fact, however, such wedges or "slivers" can become relevant whenever something akin to the extrinsic curvature of a null boundary plays a role, as it does in surface terms for the gravitational action and in connection with the Lorentzian Gauss-Bonnet theorem. Let us therefore examine the question more closely.

As earlier, let $W$ be a wedge in quadrant I with edges $a$ (spacelike) and $n$ (null), $n$ being adjacent to quadrant II. Let $N$ be a null sliver, conceived of as a very thin wedge with null edges "marked" by $n_1$ and $n_2$. There are then two ways in which we could sew $W$ to $N$, depending on whether we attach its $n$-edge to the $n_1$ edge of $N$ or to the $n_2$ edge. (See figure 10.) And unlike all the cases we have considered up until now where there was such a choice, this choice makes a difference.

Suppose we glue the $n_1$ edge of $N$ to $W$, matching $n_1$ with $n$. The combined wedge will then have edges $n_2$ and $a$, with $n_1$ lying "between" $n_2$ and $a$. Additivity in this situation would require $\theta(a, n_2) = \theta(a, n_1) + \theta(n_1, n_2)$. Substituting the known values for $\theta(a, n_1)$ and $\theta(a, n_2)$ into this equality and cancelling equal terms from the right- and left-hand sides of the equation yields $\log(a \cdot n_2) = \log(a \cdot n_1) + \theta(n_1, n_2)$. Hence

$$\theta(n_1, n_2) = \log \frac{a \cdot n_2}{a \cdot n_1} = \log(n_2 : n_1) \ , \tag{25}$$

where $n_2 : n_1 = \lambda$ if $n_2 = \lambda n_1$. But had we sewn the sliver in the other way, matching its $n_2$ edge with $W$, we would have obtained instead an angle of $\theta = \log(n_1 : n_2) = -\log(n_2 : n_1)$. The upshot is, as we anticipated, that the sign of the angle between $n_1$ and $n_2$ (i.e. that of the "opening angle" of $N$) remains indefinite until one specifies how $N$ lies on the plane next to $W$. If you flip it over, the sign of the angle also flips.

This looks rather confusing, but it can seemingly be encapsulated in a relatively simple rule: $\theta = \log(n_2 : n_1)$ if the $n_1$ edge is the one that faces "downward" toward quadrant I. Another way to say this rule is that $\theta$ is positive when the edge with the shorter marking faces downward and negative when it faces upward. (Stated like this, the rule assumes that the null vectors point upward. More generally, we would replace "facing downward" with "facing toward the spacelike quadrant of the plane adjacent to $N$".)⋆

Along with the case of a single "sliver", corresponding to a single pair of parallel null vectors, our rule generalizes naturally to the case of multiple slivers, corresponding to a succession of null vectors. If, as in figure 10, the null vectors point toward the future, and if we number them so that, proceeding from past to future, the slivers are delineated by the pairs $(n_1, n_2)$, $(n_2, n_3)$, $(n_3, n_4)$, etc, and if we take $\theta(n_j, n_{j+1}) = \log(n_{j+1} : n_j)$, then it's evident from (25) that the angles will add up correctly.

---

⋆ Compare the remarks following equation (24)

## 8. Opening angle of an arbitrary wedge in $\mathbb{M}^2$

We have now analysed enough special cases that the opening angle of any wedge whatsoever can be deduced straightforwardly from angles we already know. It suffices to subdivide the given wedge $W$ into sub-wedges $W_k$ which are narrow enough that a formula from one of the previous sections will furnish $\theta(W_k)$. The opening angle of $W$ is then simply $\theta(W) = \sum_k \theta(W_k)$. We have already seen this procedure at work in Section 6, and by the same method we could derive an explicit formula for any other case of interest. In fact there are really only five primitive cases, from which all others can be deduced by summation, namely those treated in Sections 2, 4, 5, and 7.

This seems a good place to point out how the concept of "opening angle of a wedge" differs from that of "angle between two vectors". The two are closely related of course, but the key difference which explains why we have worked mainly with the former concept is that it contains information that the latter lacks. In the Euclidean context for example, consider two orthogonal vectors, $a$ and $b$, and ask what angle $\theta$ they subtend. Is it 90 degrees or 270 or any one of an infinite number of other possibilities? All we really know is that its cosine vanishes. The question cannot be answered starting solely from the vectors themselves, but it can be answered if we associate each vector with an edge of a specified wedge $W$, i.e. if we specify how to fill in the space between $a$ and $b$ (or equivalently if we give a path that connects $a$ to $b$ while remaining within $W$).

This is also a good place to call attention to the fact that the opening angles we have defined pay no attention to any orientation that a wedge might or might not carry. As one sees clearly from the sign-rules exposed above in Sections 2–8, the positive or negative sign that a Lorentzian angle like $\theta(W)$ carries stems from the distinction between spacelike and timelike directions; it has nothing to do with any orientation of $W$ or of the vector-space in which it resides. Thanks to this independence of orientation, the Lorentzian Gauss-Bonnet theorem we will prove in the next section will encounter no difficulty in unorientable spaces like $\mathbb{R}P^2$.

## 9. Applications and Implications

*Regge Action (dihedral angle, boundary term, additivity, continuum limit)*

The most direct application of our formulas, one to which we've referred repeatedly, concerns the calculation of the gravitational action $S$ in a spacetime presented as a piecewise flat simplicial complex $\Sigma$ ($S$ being called in this setting the Regge-action). As described early on in this paper, each interior 2-simplex $\sigma$ within $\Sigma$ contributes to $S$ the product of its area $A$ by a *defect-angle* $\theta$ which is computed by adding up the dihedral angles $\theta_j$ formed at $\sigma$ by the 4-simplices which meet at $\sigma$ (the 4-simplices of its so-called "star", $\sigma^\star$) and subtracting the result from the corresponding result for flat spacetime:

$$\theta = \text{flat-value} - \sum \theta_j \qquad (26)$$

(See [6][2][3].)

Of course a dihedral angle formed by a pair of 3-simplexes (tetrahedra), is not immediately the same thing as a wedge in $\mathbb{M}^2$. That the discussion in this paper nevertheless allows us to define and evaluate the $\theta_j$ becomes clear when one realizes that the dihedral angle to which $\theta_j$ corresponds lives in effect in a quotient space of dimension two. Thus, let $\rho$ be the $j^{th}$ 4-simplex in $\sigma^\star$ and let the two faces of $\rho$ that meet at $\sigma$ be $\varphi'$ and $\varphi''$ (each being a 3-simplex). If we project $\sigma$ to zero, $\rho$ projects down to a 2-simplex (a triangle), $\varphi'$ and $\varphi''$ project down to a pair of edges of the triangle, and $\sigma$ itself projects down to the vertex at which these two edges meet.[†] (See figure 11.) These edges in turn can be identified with the vectors, $a$, $b$, $n$, etc, which feature in formulas like (18), (19), and (23) above, and the dihedral angle $\theta_j$ is then nothing but the opening angle of the triangle at (the projection of) $\sigma$. For clarity, I have described the projection in $3+1$ dimensions, but the conclusion is valid in general: $\sigma$ will always be of codimension 2, and $\rho$ will project down to a triangular wedge in $\mathbb{M}^2$ whose opening angle will furnish $\theta_j$.

So far, we have not said whether the so-called "hinge simplex" $\sigma$ is spacelike, timelike, or null. When it is spacelike, the two dimensions that get lost when $\sigma$ is projected away will also be spacelike, and our triangular wedge will live in a quotient space which is Lorentzian and isomorphic to $\mathbb{M}^2$. (Think of ignoring Cartesian coordinates $x^2$ and $x^3$ in Minkowski space.) We are in this case brought back to the situation studied in earlier sections of this paper, and all of our formulas derived there are applicable. As we have seen the total angle surrounding a point in $\mathbb{M}^2$ equals $-2\pi i$ with the conventions we have adopted. For the defect angle of a spacelike hinge $\sigma$, we thus obtain from (26),

$$\theta = -2\pi i - \sum \theta_j \qquad (27)$$

---

[†] We can regard $\rho$ as a subset of the vector-space $\mathbb{M}^4$, and then the projection in question collapses $\mathbb{M}^4$ down to $\mathbb{M}^4/V$, where $V$ is the subspace of $\mathbb{M}^4$ spanned by $\sigma$.

(Notice incidentally that in this case of a spacelike $\sigma$, an edge of our wedge is spacelike, timelike, or null precisely when the 3-simplex, $\varphi'$ or $\varphi''$, of which it is the projection is spacelike, timelike, or null.)

The hinge-simplex $\sigma$ can also be timelike, in which case the quotient space is isomorphic to the Euclidean plane. Defining the opening angles $\theta_j$ and the resulting defect-angle at $\sigma$ is then routine and presents no difficulty. The contribution of $\sigma$ to the Regge-action is once again in this case $\theta A$, the area $A$ of $\sigma$ being taken by definition to be a non-negative real number, and the defect-angle being given by

$$\theta = 2\pi - \sum \theta_j \qquad (28)$$

There is also a third possibility, which in some ways is the most interesting mathematically: $\sigma$ could be null. In this case the geometry of the quotient space falls somewhere between Lorentzian and Euclidean, and seems to be characterized by a degenerate contravariant "metric" $h^{ab}$. It seems that opening angles are not well-defined in such a space, but that for certain pairs of wedges, ratios of opening angles remain meaningful. It also seems, however, that to the extent a defect angle can be defined at all, it must vanish (see also [5]). This third case could bear further analysis, but as far as the Regge-action is concerned, the question is moot. A null hinge does not contribute to $S$, since its area is by definition zero. [3]

Along with a "bulk" contribution, the gravitational action-functional contains also a boundary term which in a simplicial manifold takes almost the same form as the bulk term. As mentioned earlier, it is a sum of terms $A\theta$, one for each codimension-two simplex $\sigma$ that lies on $\partial\Sigma$. Instead of (26) however, one now defines $\theta$ by

$$\theta = \text{flat-half-value} \ - \ \sum \theta_j \ , \tag{29}$$

where flat-half-value denotes $-i\pi$ when $\sigma$ is spacelike and $\pi$ when $\sigma$ is timelike.

The significance of the change from "flat-value" in (26) to "flat-half-value" in (29), i.e. from $-2\pi i$ or $2\pi$ to $-i\pi$ or $\pi$, emerges when one imagines $\Sigma$ as part of a larger spacetime which contains $\Sigma$ in its interior. Let this larger spacetime be $\Sigma \cup \Sigma'$, where $\Sigma \cap \Sigma' = \partial\Sigma$. From the definitions, (26) and (29), it is obvious that in this situation

$$S(\Sigma \cup \Sigma') = S(\Sigma) + S(\Sigma') \tag{30}$$

because the two boundary terms (29) coming from $\Sigma$ and $\Sigma'$ combine to give the single bulk term (26) for $\Sigma \cup \Sigma'$. Conversely, this additivity is ultimately the raison d'etre for the boundary term (29) and the explanation of its particular form. If you start with (26) for the "interior" hinges, and you ask yourself, What action could I attribute to the boundary hinges so that the full action will be additive?, you will be led inevitably to (29) as the obvious answer.

REMARK Equation (30) will fail if $\Sigma$ and $\Sigma'$ share a boundary simplex $\sigma$ that remains on the boundary of their union. In order for it to hold true, one needs that $\partial\Sigma \cap \partial\Sigma'$ be disjoint from $\partial(\Sigma \cup \Sigma')$. In a similar way, it will fail in general when the larger manifold is the union of three or more pieces. [7], [8]

The action-additivity expressed implicitly by (29) and explictly by (30) has two important implications for the continuum theory. First of all, (29) implies the existence of so-called *corner terms* in the action $S$, whose form it also furnishes. Second of all, (30) implies that when all boundary terms are included, the total action $S(\Sigma)$ will be stationary under small variations about a solution that holds fixed the induced metric of $\partial\Sigma$, an implication we might summarize by saying, *angle-additivity implies action-stationarity*. Let us take these implications in turn.

The "boundary defect angle" $\theta$ of (29) is *already* a corner-term for the simplicial spacetime $\Sigma$, being supported on the codimension-two simplices where one boundary 3-simplex meets another. In a limit where the simplices of $\Sigma$ become infinitely fine in such a way as to converge to a smooth manifold $M$ with corners, the term (29) will remain as a corner term where $M$ has corners, while it will converge to some smooth boundary term on the rest of $\partial M$. Accordingly, the surviving corner terms (per unit area) will be determined directly in terms of the opening angles we have derived in this paper. This chain of reasoning, if followed through for all the various types of corners, should provide a complete explanation of, and a recipe for, the corner terms found in [9] and [5] including especially the novel type of corner where a null boundary-segment meets another null or non-null portion of the boundary. (It would make a good project to confirm this claim in detail!)

It is also very plausible that the smoother boundary terms, i.e. those that do not pertain to corners, could be understood in the same way. In fact it is not hard to verify (cf. [7] [10]) that at a hinge-simplex $\sigma$ where two simplices belonging to a spacelike portion of $\partial\Sigma$ meet, the trace of the extrinsic curvature takes the form of a $\delta$-function supported on $\sigma$ (the 3-simplexes themselves being internally flat), which when integrated reproduces (29). It would be interesting to try to derive the well-known $\text{Tr}\,K$ boundary-term rigorously from these observations, and even more interesting to understand similarly the "surface gravity" contributions from the null portions of $\partial M$, starting from equations (24) and (25).

In those cases where a null boundary is involved, the ambiguity in parameterizing its null generators should correspond to the "marking ambiguity" that arises when one wishes to compute the opening angle of a wedge, either of whose edges is lightlike.

### additivity and extremality

The second thing that needs fleshing out is my claim above that the additivity of $S$ implies its extremality. In fact this is easily demonstrated by embedding $M$ into a larger spacetime $M \cup M'$ as we did above simplicially with $\Sigma$. Let this be done and let the metric of $M \cup M'$ solve the Einstein equations. Then by definition, $S(M \cup M')$ will be stationary under small variations of the metric. But by hypothesis (i.e. with the appropriate boundary terms included) we also know that $S(M \cup M') = S(M) + S(M')$, whether or not we are at a solution. Now restrict the variation so that the metric on $M'$ remains unchanged, whence $\delta S(M') = 0$. The metric on $M$ can still vary freely, except that for the sake of continuity with $M'$, the induced metric on $\partial M$ will also have to remain unchanged.[♭] Under these conditions, $0 = \delta S(M \cup M') = \delta S(M)$, and we have proven that $\delta S(M) = 0$ under arbitrary variations which fix the induced metric on $\partial M$.

REMARK  In seeking a rationale for choosing boundary terms in the gravitational (or any other) action, one often invokes the requirement that the action be additive as in (30) when one builds a spacetime from two pieces whose induced boundary-metrics agree (which in turn is tantamount to asking that the action be the integral of a Lagrangian density that is of first differential order). One also often asks that the action be stationary under variations that preserve the boundary-geometry. We have seen here how closely related these two conditions are.

## an unfamiliar familiar fact about triangles

An amusingly familiar, but also very useful fact that follows directly from our definitions is the Lorentzian counterpart of the Euclidean theorem that the interior angles of a triangle add up to $\pi$. In order to get some practice in Lorentzian Trigonometry in the spirit of [11], let us prove this counterpart in $\mathbb{M}^2$, paying special attention to the possibility that some of the edges can be lightlike. To help bring out the parallelism between the Euclidean and Lorentzian cases, I will in the following let $h$ be the value of a "straight angle" in the respective cases: $h = \pi$ if Euclidean and $h = -i\pi$ if Lorentzian. Our goal is then to prove that the angles of a triangle sum to $h$.

To set up the proof, start with a triangle with vertices $A$, $B$, $C$, and extend the edges as indicated in figure 12, so that $A = (C + C')/2$ is midway between $C$ and $C'$. In terms of vectors, we have then $[AC] = -[AC']$, or $[CA] = [AC']$, where $[XY]$ stands for the vector from $X$ to $Y$. Similarly $[AB] = [BA']$ and $[BC] = [CB']$, Now consider the situation at

---

[♭] One might think that continuity would also force the transverse derivatives of the metric to be unvaried. To analyze this issue fully, one needs to be careful about defining the differentiable structures of $M$, $M'$ and $M \cup M'$ and tracing out the consequences of diffeomorphism-invariance, but suffice it to say that in the end there is no such restriction. In particular the extrinsic curvature can vary freely, as one sees with particular clarity in the simplicial context.

vertex $A$. The angle of interest, indicated by $\alpha$ in the diagram, is the opening angle of the wedge $BAC$. By additivity, this angle is related to $\alpha'$ by

$$\alpha + \alpha' = h, \tag{31}$$

from which we conclude, by adding this to the analogous equations for the other two vertices, that

$$(\alpha + \beta + \gamma) + (\alpha' + \beta' + \gamma') = 3h \tag{32}$$

But we also know that $\alpha' = \theta([AC'], [AB]) = \theta([CA], [AB])$, which when added to its counterparts at $B$ and $C$ becomes

$$\alpha' + \beta' + \gamma' = \theta([AB], [BC]) + \theta([BC], [CA]) + \theta([CA], [AB]) \tag{33}$$

This last expression, being the angle accrued in going full circle from $[AB]$ back to $[AB]$ (or equivalently the net opening angle of three wedges that fit together with neither gap nor overlap), is plainly $2h$ (namely $2\pi$ or $-2\pi i$, respectively). Combining (32) with (33), produces finally

$$\alpha + \beta + \gamma = h \tag{34}$$

This completes the proof, but where were the subtleties involving null vectors hiding? First of all in (31), which when $AC$ is lightlike, relies on (24), which in turn holds only because $-[AC] : [AC'] = 1$ (and of course because opening angles are additive even when null edges are involved). And secondly in the conclusion that the right hand side of (33) is $2h$, which holds only because, the vector $[AB]$ (for example) is not only the same ray in both its occurrences in (33) but literally the same vector. In effect we have construed the angle $BAC$ as a *marked* wedge by using the two other vertices, $B$ and $C$, as marks. In a simplicial complex, consequently, every wedge that occurs is automatically marked. So there were subtleties, but if we avert our gaze from them, we will see no difference between the Euclidean and Lorentzian proofs.

REMARK The "complementary angle" $\alpha'$ at vertex $A$ (which in another context is equal to the exterior defect-angle) can also be interpreted as the angle between the normal vectors to sides $AB$ and $AC$ of the triangle, but one would have to be careful about signs. If we chose the normals to be *outward* when interpreted as covectors, and if the resulting sign-rules could be sorted out, we might obtain an alternative, and even slightly simpler, way to deduce that $\alpha' + \beta' + \gamma' = 2h$.

*Lorentzian Gauss-Bonnet theorem for any topology*

By taking advantage of the triangle-theorem just demonstrated, we can, simply by mimicking the analogous deductions for Euclidean signature, prove with almost no extra effort, a simplicial Lorentzian Gauss-Bonnet theorem! To that end, let $\Sigma$ be a two-dimensional (Lorentzian) simplicial complex as above, which we can take for definiteness to be the triangulation of a two-dimensional manifold (not necessarily orientable and possibly with boundary). Let $\chi$ be the Euler number of $\Sigma$, and let $S$ be its Regge-action as defined earlier.[*] We want to show that the action is nothing but the Euler number, or more precisely that

$$S = -2\pi i \chi \tag{35}$$

which we can write as $S = 2h\chi$ if, as before, we define $h$ to be $-i\pi$. Also write $V$ for the number of vertices (0-simplices) of $\Sigma$, $E$ for the number of edges (1-simplices), and $F$ for the number of triangles (2-simplices). By definition, $\chi = V - E + F$.

We will need to distinguish "interior" simplices from "boundary" simplices, the latter being simplices which are subsets of $\partial \Sigma$. Let $V^o$ denote the number of interior vertices, and $V^\partial$ the number of boundary vertices, and similarly for $E^o$ and $E^\partial$. We will establish the following three equations, which together will yield the desired equality.

$$S/h = 2V^o + V^\partial - F \tag{36}$$

$$-2E^o - E^\partial + 3F = 0 \tag{37}$$

$$V^\partial - E^\partial = 0 \tag{38}$$

The first equation (36) comes from summing (27) and (29) over all the 0-simplices of $\Sigma$. In two dimensions, a hinge-simplex, being of codimension two, is simply a 0-simplex or vertex, and its area is $A = 1$ by convention. Hence, the total action $S$ is nothing but the sum over all vertices of their defect-angles, as given by either (27) for interior vertices or (29) for boundary vertices. In this sum, each interior angle of each triangle appears exactly once, and since the sum of the angles for any given triangle is $h$, the contribution of the $\theta_j$ terms in (27) and (29) to $S/h$ is simply $-Fh/h = -F$ where $F$ is the total number of triangles. This explains the last term in (36). In addition, the constant terms of $2h$ in (27) and $h$ in (29) contribute $2hV^o + hV^\partial$ to $S$. This explains the first two terms in (36). The second equation (37) reflects the facts that each interior edge is incident on exactly two triangles, while each boundary edge is incident on exactly one, while each triangle is incident on exactly three edges; therefore $2E^o + 1E^\partial = 3F$. The third equation (38) records that, because the boundary is simply a cycle (or a disjoint union of cycles) of the form, vertex-edge-vertex-edge-vertex-etc, there are exactly as many boundary vertices as boundary edges. Finally, equation (35) results if one adds the left-hand sides of (37) and (38) to the right-hand side of (36). QED.

---

[*] The normalization is that of $(1/2) \int R dV$

The Gauss-Bonnet theorem we have just proven presupposes remarkably little about the topology of $\Sigma$. Let us apply it for example to the "trousers cobordism" $M$ that mediates the splitting of one circle into two. We can assume that the metric on $M$ vanishes at a single "Morse point", but is invertible everywhere else. If our theorem (35) persists in the continuum limit sketched above, it will imply immediately (given that the trousers is homeomorphic to a sphere with three disks removed, and therefore has Euler number, $\chi = 2 - 3 = -1$) that $S = +2\pi i$. We can even do better than this, because one can furnish $M$ with a metric which is flat everywhere except at the Morse-point (see e.g. [12]), and for such a metric the simplicial approximation is already exact, so that our theorem applies as is. Now in the quantum-gravity amplitude $e^{iS}$, an action of $2\pi i$ yields a damping factor of $e^{-2\pi}$, in agreement with what was found in [1] by another method. Of course the agreement in sign (damping vs. enhancement) is not entirely accidental. It stems from the choice we made in analytically continuing (6) past the branch point at $b \cdot b = 0$ to obtain (8), and our choice involved a closely related kind of complexified metric to that employed in [1].

In any case, we now have a convenient mnemonic to help remember the choice of sign made herein. It is the one for which the trousers-cobordism is dynamically suppressed, while the "yarmulke-cobordism" is enhanced. For the yarmulke there's no null direction at all at the Morse-point, whence we get $-2\pi i$ from (27); for the trousers there are eight of them, whence we get $-2\pi i - 8 \times -i\pi/2 = +2\pi i$.

REMARK An interesting question is where (apart from any Morse-points that might be present) an imaginary action like that of (35) could come from in a continuum calculation. In part it could come from the corner terms, but what if the boundary were entirely smooth? In that situation it would seem to have to arise where the tangent vector passed from spacelike or timelike to null, but how to define the boundary integrand there? Or is such a boundary inadmissible in a Lorentzian spacetime?

### *final comments*

One implication of our work in this paper is that the seemingly natural proposal to define angles involving lightlike directions by first "regulating" the metric and then taking some sort of renormalized limit seems to be untenable. The problem is that even if, for example, we add a positive-definite imaginary part to the metric of $\mathbb{M}^2$ so that the denominator of (6) no longer vanishes when $a$ or $b$ becomes lightlike, we will still end up with a concept of angle that refers to unmarked rays. Only if the renormalization somehow smuggled in "marking" could such an approach succeed. Nevertheless, deforming the metric into the complex *did* prove helpful in connection with our analytic continuation in Section 3.

Finally, a few comments of a general nature on what was done above. A surprising feature of the expressions we have derived is that angles involving null vectors are easier to write down than angles involving only spacelike or timelike vectors (likewise for the corresponding corner terms in the gravitational action). On the other hand, in order to be dimensionally correct we had to introduce a reference length $l_0$ into our definitions of certain angles involving null vectors, and the physical significance of this length, if any, is not evident. For angles not confined to a single (open) quadrant, we also had to make a largely arbitrary choice of where to put the imaginary contributions of $-i\pi/4$. But perhaps the biggest (and certainly most welcome) surprise is that, once the appropriate definitions and conventions are in place, Lorentzian trigonometry appears almost indistinguishable from its Euclidean cousin.

It's a pleasure to thank Sumati Surya and Fay Dowker for comments on these topics, and special thanks go to Safiya Sivjee for converting my scribbled diagrams into the beautiful figures that the paper now enjoys. This research was supported in part by NSERC through grant RGPIN-418709-2012. This research was supported in part by Perimeter Institute for Theoretical Physics. Research at Perimeter Institute is supported by the Government of Canada through Industry Canada and by the Province of Ontario through the Ministry of Economic Development and Innovation.

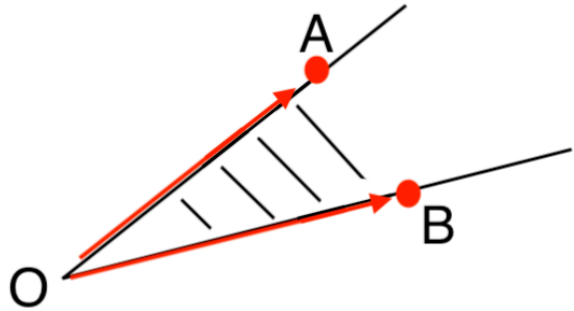

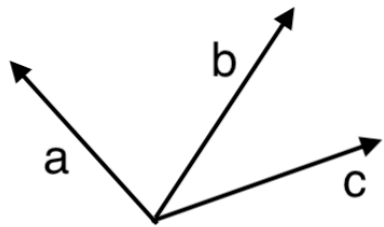

*Figures 1 and 2.*
1. A wedge *marked* by the vectors $[OA]$ and $[OB]$.
2. Three vectors with $b$ between $a$ and $c$

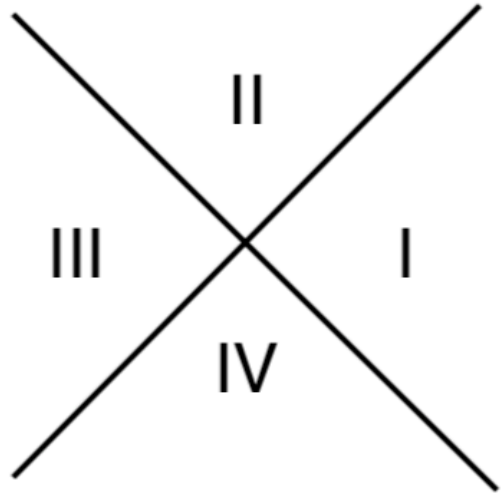
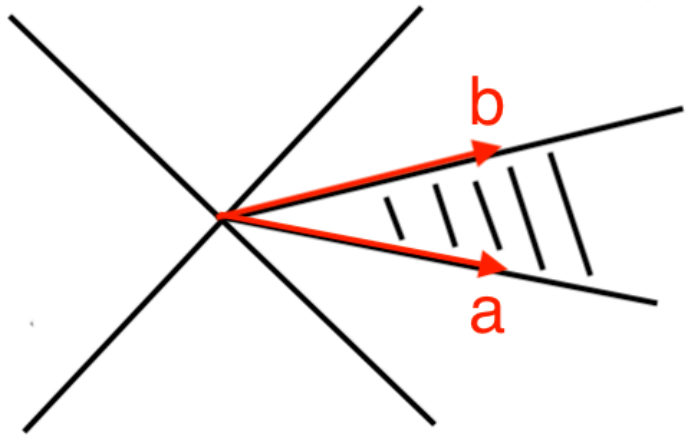

Figure 3. The four quadrants and a wedge in quadrant I

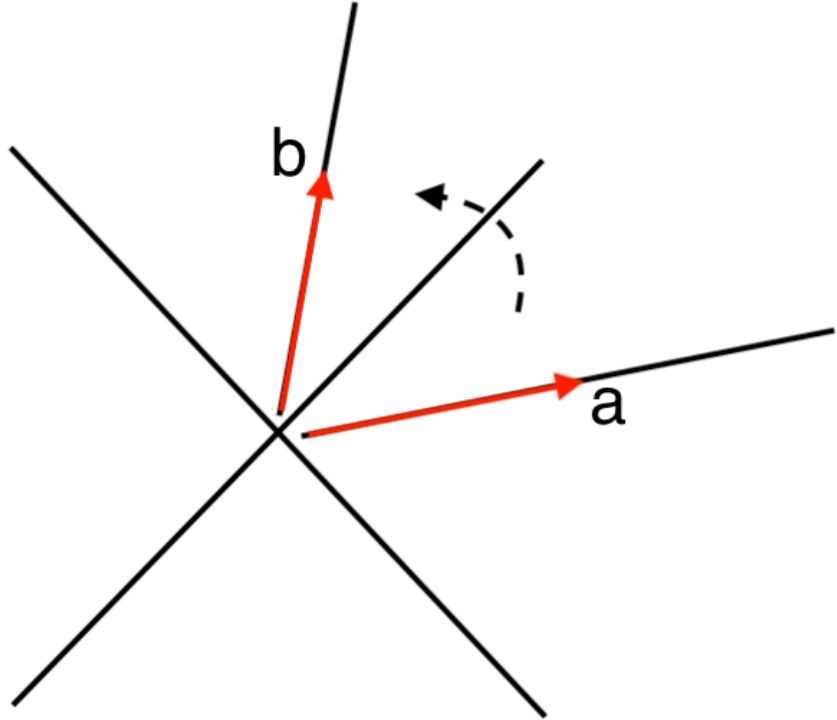

*Figure 4.* A wedge $W$ spanning two quadrants. The edge marked by $a$ is in quadrant I, the $b$-edge is in quadrant II. We analytically continue the upper edge of $W$ from $a$ to $b$

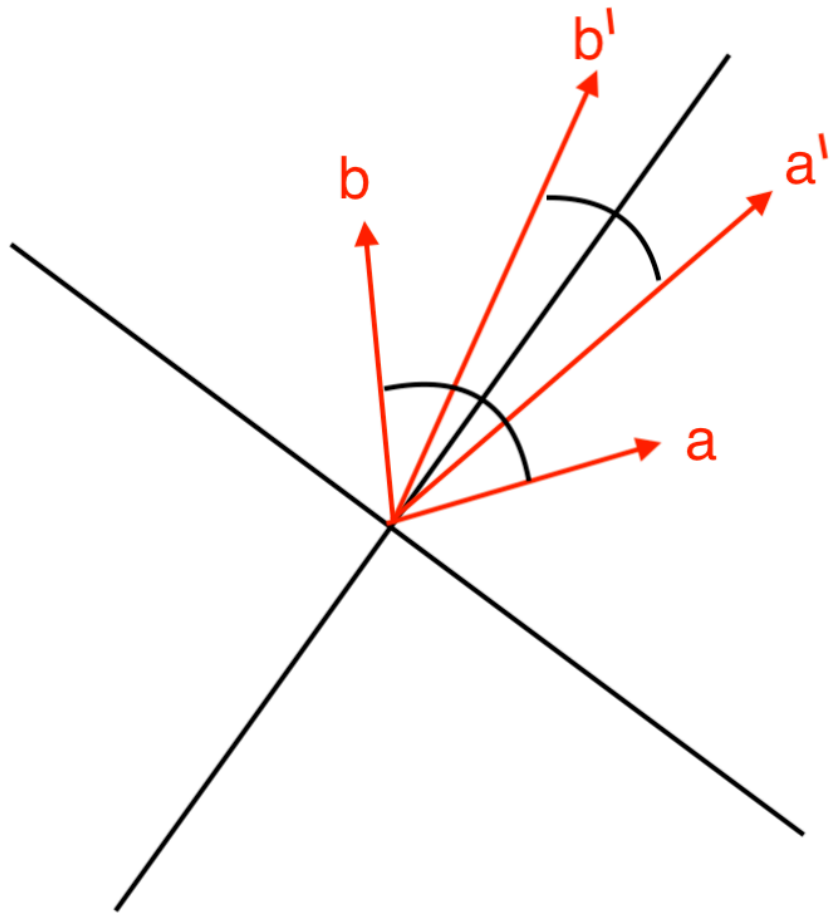

*Figure 5.* Illustrating why timelike angles have to be negative if spacelike ones are positive

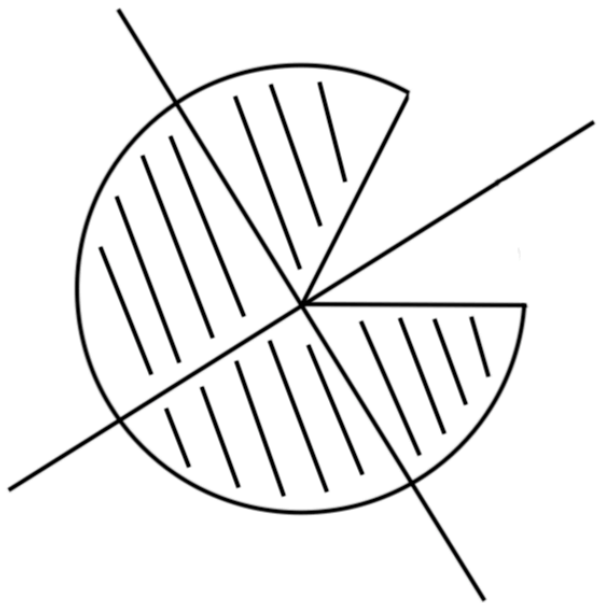

*Figure 6.* A wedge need not be convex.

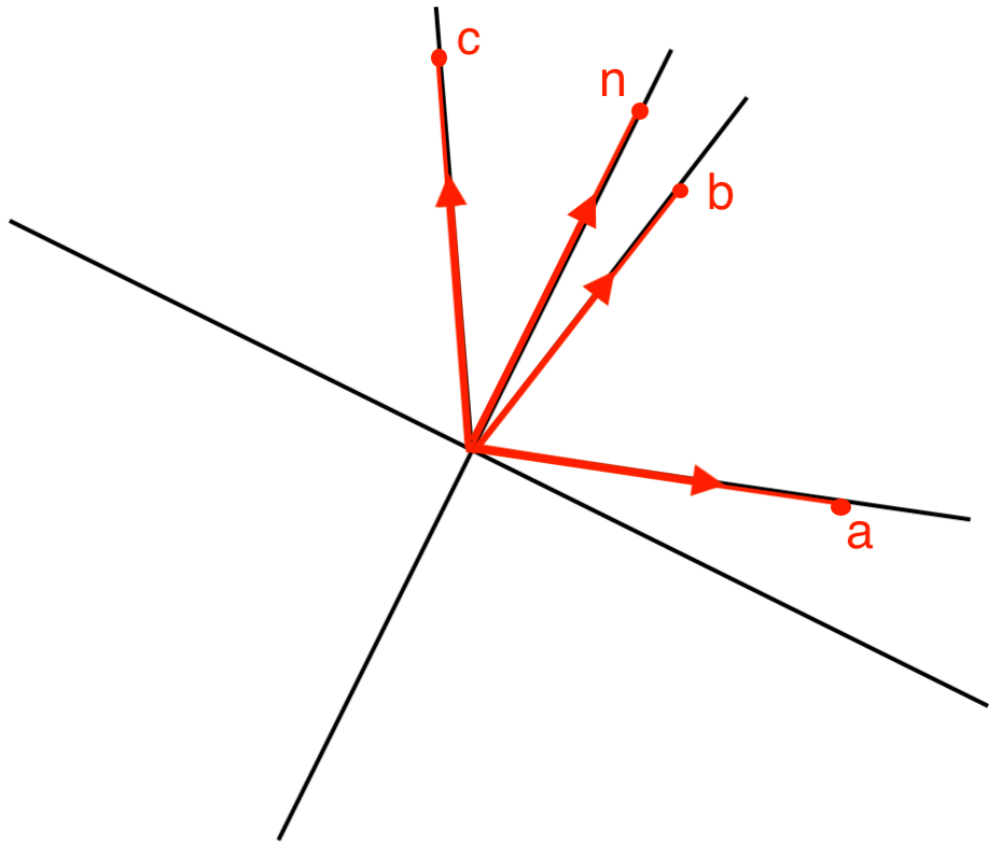

*Figure 7.* The vectors that enter the derivation of (18) and (19).

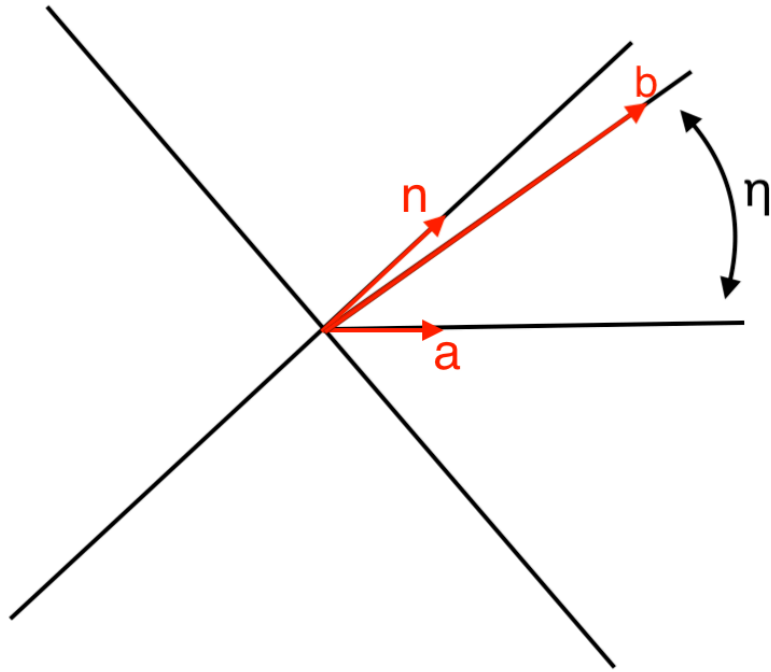

*Figure 8.* The boosted vector, *b*, approaches the light-cone as "closely" as desired

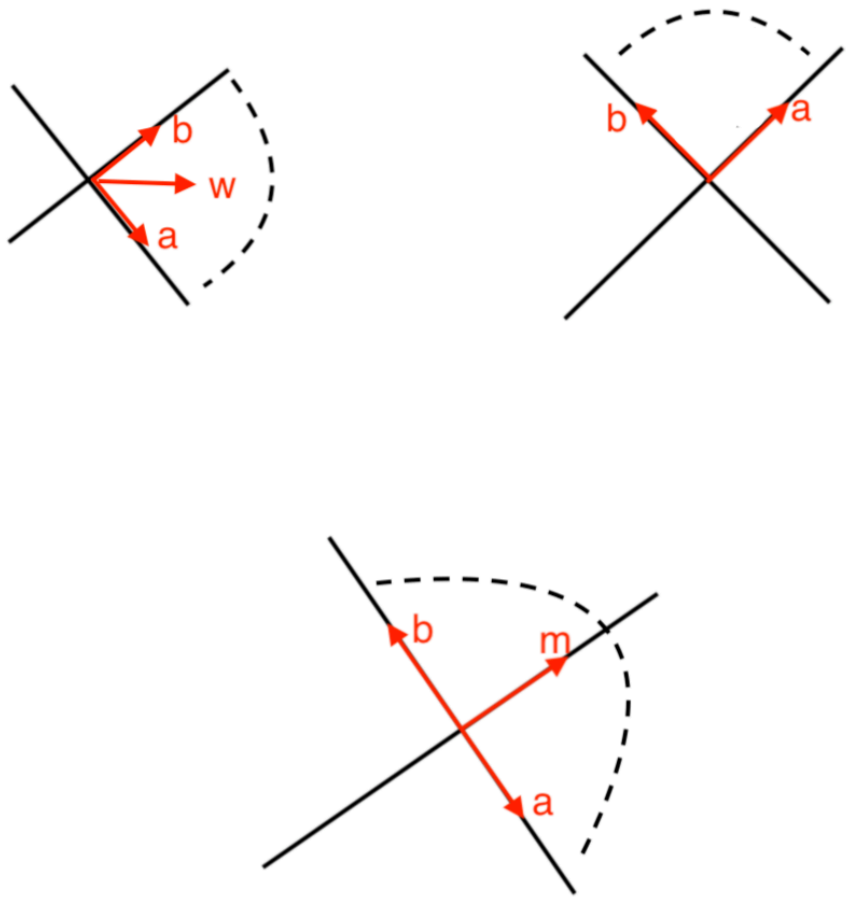

*Figure 9.* The opening angle of a wedge with two lightlike edges: three sub-cases

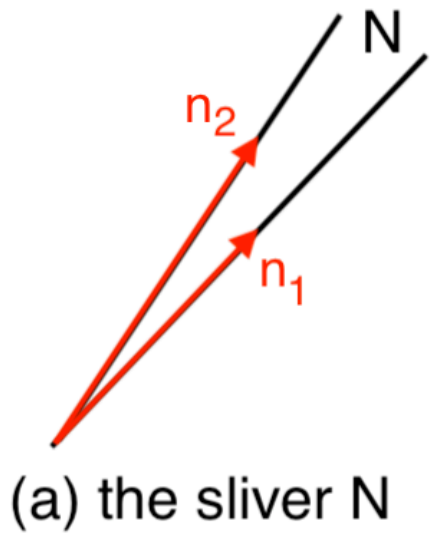

(a) the sliver N

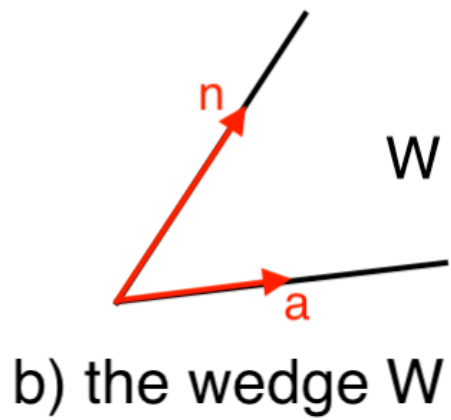

b) the wedge W

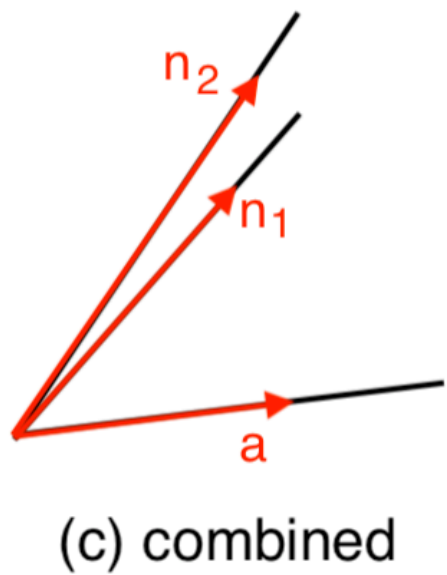

(c) combined

*Figure 10.* Two ways to glue an "infinitesimal wedge" or "sliver" to the wedge $W$. (a) the sliver. (b) the wedge $W$. (c) the combined wedge if $n_1$ rather than $n_2$ is matched with $n$.

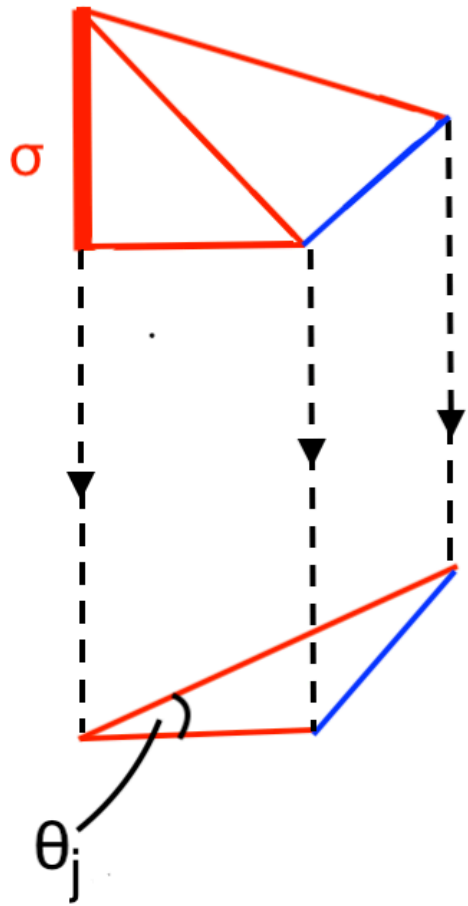

*Figure 11.* The projection that converts a dihedral angle to a wedge in $\mathbb{R}^2$ is illustrated in $2+1$ dimensions. The hinge simplex $\sigma$ collapses to a single point, preserving the opening angle $\theta_j$.

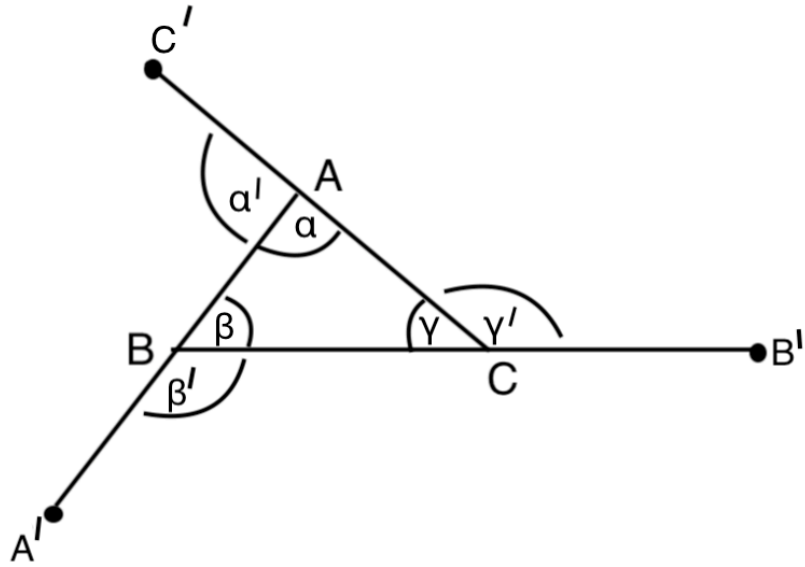

*Figure 12.* Whether a triangle is Lorentzian or Euclidean, its internal angles sum to a straight angle.